\documentclass[final,3p,times,authoryear]{elsarticle}

\usepackage{amssymb}
\usepackage{amsmath}

\usepackage{listings}
\usepackage{enumerate}
\usepackage{color}

\newcommand{\eq}[1]{Eq.\ \ref{#1}}
\newcommand{\eqs}[2]{Eqs.\ (\ref{#1})-(\ref{#2})}
\newcommand{\Fig}[1]{Fig.\ \ref{#1}}
\newcommand{\Listing}[1]{Listing \ref{#1}}
\newcommand{\Sec}[1]{Sec.\ \ref{#1}}
\newcommand{\mytable}[1]{Table\ \ref{#1}}
\newcommand{\bld}[1]{{\bf {#1}}}
\newcommand{\tim}[2]{t_{\rm {#1}}^{\rm {#2}}}
\def\cf{{\it cf.}}
\def\eg{{\it e.g.}}

\def\m11{{\rm\bf M11}}

\journal{Parallel Computing}

\begin{document}

\lstset{language=Fortran,basicstyle=\footnotesize}

\begin{frontmatter}



\title{GPU parallelization of a hybrid pseudospectral fluid turbulence
  framework using CUDA}


\author[dradd1]{Duane Rosenberg}
\address[dradd1]{1401 Bradley Drive, Boulder, CO 80305}

\author[pmadd1]{Pablo D. Mininni}
\address[pmadd1]{Departamento de Física, Facultad de Ciencias Exactas
  y Naturales \& IFIBA, CONICET, Ciudad Universitaria, 1428 Buenos
  Aires, Argentina}

\author[rradd1]{Raghu Reddy}
\address[rradd1]{CSRA Inc., at NOAA/NWS/NCEP/Environmental Modeling
  Center, 5830 University Research Court, Suite 2146 College Park, MD
  20740 USA}

\author[apadd1,apadd2]{Annick Pouquet}
\address[apadd1]{National Center for Atmospheric Research, P.O. Box 3000, Boulder,CO80307}
\address[apadd2]{Laboratory for Atmospheric and Space Physics, CU, 
  Boulder, CO 80309-256 USA}

\begin{abstract}
An existing hybrid MPI-OpenMP scheme is augmented with a CUDA-based
fine grain parallelization approach for multidimensional distributed
Fourier transforms, in a well-characterized pseudospectral fluid
turbulence code. Basics of the hybrid scheme are reviewed, and 
heuristics provided to show a potential benefit of the CUDA
implementation. The method draws heavily on the CUDA runtime library
to handle memory management, and on the cuFFT library for computing
local FFTs. The manner in which the interfaces are constructed to
these libraries, and ISO bindings utilized to facilitate platform
portability, are discussed. CUDA streams are implemented to overlap
data transfer with cuFFT computation. Testing with a baseline solver
demonstrates significant aggregate speed-up over the hybrid MPI-OpenMP
solver by offloading to GPUs on an NVLink-based test system. While
the batch streamed approach provides little benefit with NVLink, we
see a performance gain of $30\%$  when tuned for the optimal number
of streams on a PCIe-based system. It is found that strong GPU
scaling is ideal, or slightly better than ideal, in all cases. In
addition to speed-up measurements for the fiducial solver, we also
consider several other solvers with different numbers of transform
operations and find that aggregate speed-ups are nearly constant for
all solvers.
\end{abstract}

\begin{keyword}
Computational fluids \sep Numerical simulation \sep
MPI \sep OpenMP \sep CUDA \sep Parallel scalability
\end{keyword}

\end{frontmatter}


\section{Introduction}
\label{sec:intro}

Turbulent flows and multi-scale interactions are often studied
computationally using the pseudospectral numerical method
\citep{orszag1972,canuto1988}. This is because of its inherent high
order truncation, its consequent lack of diffusivity and dispersion
and, importantly, because of its local (per node) computational
complexity, which, using fast spectral transforms at a linear grid
resolution of $N$, goes as $N \log N$ instead of $N^2$. The grid
resolutions required to study turbulent flows--without resorting to
modeling--vary as a high power of the Reynolds number that
characterizes the turbulence. For geophysical fluids with Reynolds
numbers often larger than $10^8$, this can translate into grids with
more than $10^{18}$ gridpoints in three dimensions, yielding a truly
exascale computation. For this reason, even when lower Reynolds
numbers are considered, computational fluid dynamics (CFD) approaches
to turbulence require efficient parallelization methods with good
scalability up to very large number of processors.

In \citet{hybrid2011} (hereafter, \m11) we presented a hybrid
MPI-OpenMP pseudospectral method and showed its scalability and
parallel efficiency up to reasonably high core counts. We also
provided guidance on optimization on NUMA systems which touched on
issues of MPI task and thread affinity to avoid resource contention. 
The present paper builds upon this previous study. It was recognized 
early that we could achieve significant performance gains if we could 
essentially eliminate the cost of local Fast Fourier Transforms (FFTs)
and perhaps other computations by offloading this work to an
accelerator. The CUDA FFT library \citep{cuFFT_developer_link}
together with the CUDA runtime library \citep{cudaruntime_link} allow
us to do this. The idea is straightforward: copy the data to the
device, compute the local transform and perhaps other local
calculations, and copy back so that communication and additional
computation may be carried out.

Because of the shear number of applications and because of the desire
to reach higher Reynolds numbers and larger and more complex
computational domains, fluid and gas dynamics applications have often
been at the forefront of development for new computational
technology. The most successful of these efforts for accelerators have
generally been particle-based or particle-like methods that are known
to scale well and have been ported to GPU-based systems
\citep{ripesi2014,yokota2013}, and even to Cell processor-based 
systems \citep{sturmer2009}. It is not uncommon, however, to find even
with ostensibly highly scalable methods that reported performance
measurements are  restricted to single nodes or even to single kernels
on a single accelerator rather than to holistic performance. On the
other hand, global modeling efforts have shown superb aggregate
performance on GPU-based systems \citep{govett2017}, but such codes
often used for numerical weather or climate are typically low-order,
and not suited for studies of detailed scale interactions.

While there are some efforts in the literature to port pseudospectral
methods to GPUs \citep[see, e.g.,][]{thibault2009}, most appear to be
tailored to smaller accelerated desktop solutions that avoid the issue
of network communication that necessarily arises in massively
distributed applications of the method (see \Sec{sec:setup}). This
communication is usually considered such a severe limitation
\citep{yokota2013} that it has repeatedly sounded the death knell for
pseudospectral methods for some time, reputedly preventing scaling to
large node counts and yielding poor parallel efficiency in multi-node
CPU- and GPU-based systems. This may eventually be borne out; however,
as demonstrated in \m11 and in subsequent work \citep{rosenbergbo},
our basic hybrid scheme continues to scale on CPU-based multicore
systems with good parallel efficiency, likely due in part to the 1D
domain decomposition scheme that is used (\Sec{sec:setup}). Indeed,
other authors
\citep{dmitruk2001,kaneda,yeung2005,donzis2008,chatterjee2018} have 
also seen good scaling of pseudospectral methods using alternative
decompositions. As we will see below, good scaling is expected to
continue on emerging GPU-based systems with our new CUDA
implementation.

In this paper we present a new method based on the hybrid algorithm
originally discussed in \m11 that allows us to run on GPUs. The problem
is formulated and given context in \Sec{sec:setup}, where we also
update the formulation in \m11 to accommodate anisotropic grids. In
\Sec{sec:implement} the implementation of the method is discussed,
showing how the interfaces to the CUDA runtime and cuFFT libraries are
handled, as well as how portability is maintained. Results showing the
efficacy of the CUDA implementation on {\it total} runtimes are
provided in \Sec{sec:results} for our reference equations, and
aggregate speed-ups for other solvers are also presented. We establish
that the transfer time to and from the device becomes a new cost that
can lessen the gains achieved by computing on the GPUs, and we show
how CUDA streams may be used to diminish the impact of the transfer on
different systems. Finally, in \Sec{sec:conclusion} we offer our
conclusions, as well as some observations about the code and future
work. Overall, the method we present gives significant speed-ups on
GPUs, ideal or better than ideal parallel scaling when multiple nodes
and GPUs are used, and can be used to generate fast parallel
implementations of FFTs, or three-level (MPI-OpenMP-CUDA)
parallelizations of pseudospectral CFD codes, or of other CFD methods 
requiring all-to-all communication.

\section{Problem description}
\label{sec:setup}

Our motivation is to solve systems of partial differential equations
(PDEs) that describe fluids in periodic Cartesian domains for purposes
of investigating turbulent interactions at all resolvable scales. A
prototypical system of PDEs that describes the conservation of
momentum of an incompressible fluid in a stably stratified domain is
given by the Boussinesq equations:
\begin{eqnarray} 
\partial_t {\mathbf u}  +  \nabla p  + {\mathbf u} \cdot \nabla
     {\mathbf u}
  &=& - N_{bv} \theta \hat z +\nu \Delta {\mathbf u} , 
     \label{eq:mom} \\
\partial _t \theta +  {\mathbf u} \cdot \nabla \theta 
  &=& N_{bv} w +  \kappa \Delta \theta , \label{eq:temp} \\
\nabla \cdot {\bf u} &=& 0, \label{eq:divv}
\end{eqnarray}
in which ${\mathbf u}$ is the velocity, $p$ the pressure (effectively
a Lagrange multiplier used to satisfy the incompressibility constraint
given by \eq{eq:divv}), $\theta$ the temperature (or density)
fluctuations, and $N_{bv}$ is the Brunt-V\"ais\"al\"a 
frequency which establishes the magnitude of the background
stratification. The dissipation terms are governed by the viscosity
$\nu$, and the scalar diffusivity $\kappa$. These equations,
essentially the incompressible Navier-Stokes equations together with
an active scalar (the temperature) and additional source terms, are
relevant for studies of geophysical turbulence, and in the following will be
considered as our reference equations for most tests. 

We will also consider three other systems of PDEs often used for
turbulence investigations in different physical contexts, of similar
form except for the last case relevant for condensed matter
physics. When in the equations above the temperature is set to zero
($\theta=0$), we are only left with the incompressible Navier-Stokes
equations, given by \eq{eq:mom} and \eq{eq:divv}. This set of
equations is used to study hydrodynamic (HD) turbulent flows. In space 
physics magnetohydrodynamic (MHD) flows are often considered, which
describe the evolution of the velocity field 
${\mathbf u}$ and of a magnetic field ${\mathbf B}$:
\begin{eqnarray} 
\partial_t {\mathbf u}  +  \nabla p  + {\mathbf u} \cdot \nabla
     {\mathbf u}
  &=& {\mathbf B} \cdot \nabla {\mathbf B} + \nu \Delta 
     {\mathbf u} \label{eq:uMHD} , \\
\partial _t {\mathbf B} - \nabla \times ({\mathbf v} \times 
     {\mathbf B}) 
  &=& \eta \Delta {\mathbf B} \label{eq:bMHD} , \\
\nabla \cdot {\bf u} = \nabla \cdot {\bf B} &=& 0 ,
\end{eqnarray}
where $\eta$ is the magnetic diffusivity. Finally, to study quantum
turbulence the Gross-Pitaevskii equation (GPE) describes the evolution
of a complex wavefunction $\psi$:
\begin{equation}
i \hbar \partial_t \psi = - \hbar^2/(2m) \Delta \psi 
     + g \vert \psi \vert^2 \psi , \label{eq:gpe}
\end{equation}
where $g$ is a scattering length and $m$ a mass. The four sets of
equations (Boussinesq, HD, MHD, and GPE) allow us to explore the
performance of the method for different equations often used in CFD
applications.

Our work relies on the Geophysical High Order Suite for Turbulence
(GHOST) code (\m11),  which provides a framework for solving these
types of PDEs using a pseudospectral method
\citep{patterson1971,gottlieb1984,canuto1988}. To discretize the PDE,
in general, each component of the fields is expanded in terms of a
discrete Fourier transform of the form
\begin{equation}
\hat{\phi}_{pqr} = \sum_{i=j=k=0}^{N_x-1,N_y-1,N_z-1} \phi_{ijk}
     \exp[-2\pi i(p L_x/N_x + q L_y/N_y+ r L_z/N_z)] , 
\label{eq:fftsum}
\end{equation}
where $\hat{\phi}_{pqr}$, represent the (complex) expansion
coefficients in spectral space. We focus here on 3D transforms, and
neglect any normalization.  The expansion above represents the 
{\it forward} Fourier transform; a similar equation holds for the
{\it backward} transform but with a sign change in the exponential
(again, neglecting any normalization). The indices $(i,j,k)$ represent
the physical space grid point, and $(p,q,r)$ represent the wave number
grid location. Note that both the physical box size, $L_i$, and the
number of physical space points, $N_i$, may be specified independently
in each direction in our implementation enabling non-isotropic
expansions in the $(L_x, L_y, L_z)$ domain. This is a major difference
between the current algorithm and the version discussed in \m11 \citep{Sujovolsky2018}.

Taking the continuous Fourier transform and utilizing the discrete
expansion given by \eq{eq:fftsum} in the system of PDEs, yields a
system of time-dependent ordinary differential equations (ODEs) in
terms of the complex Fourier coefficients, $\hat{\phi}_{pqr}$, for
each field component. These represent the solution at a point (or 
{\it mode}) in the associated 3D spectral space. In GHOST, the system
of ODEs is solved using an explicit Runge-Kutta scheme (of 2nd-4th
order); explicit time stepping is used to resolve all wave modes. In
the {\it pseudospectral} method, the nonlinear terms appearing in 
\eqs{eq:mom}{eq:divv} (or in any other of the PDEs considered) are
computed in physical space, and then transformed into spectral space
using the multidimensional Fourier transform, in order to obviate the
need to compute the convolution integrals explicitly, which is
expensive. Since the pressure is a Lagrange multiplier, its action is
taken by projecting the nonlinear term in spectral space onto a 
divergence-free space, so that the velocity update will satisfy
\eq{eq:divv}. If not for the physical space computation of nonlinear
terms, the method, as is done in pure spectral codes, would solve the
PDEs entirely in spectral space.

\begin{figure}
\begin{center}
\includegraphics[height=20pc]{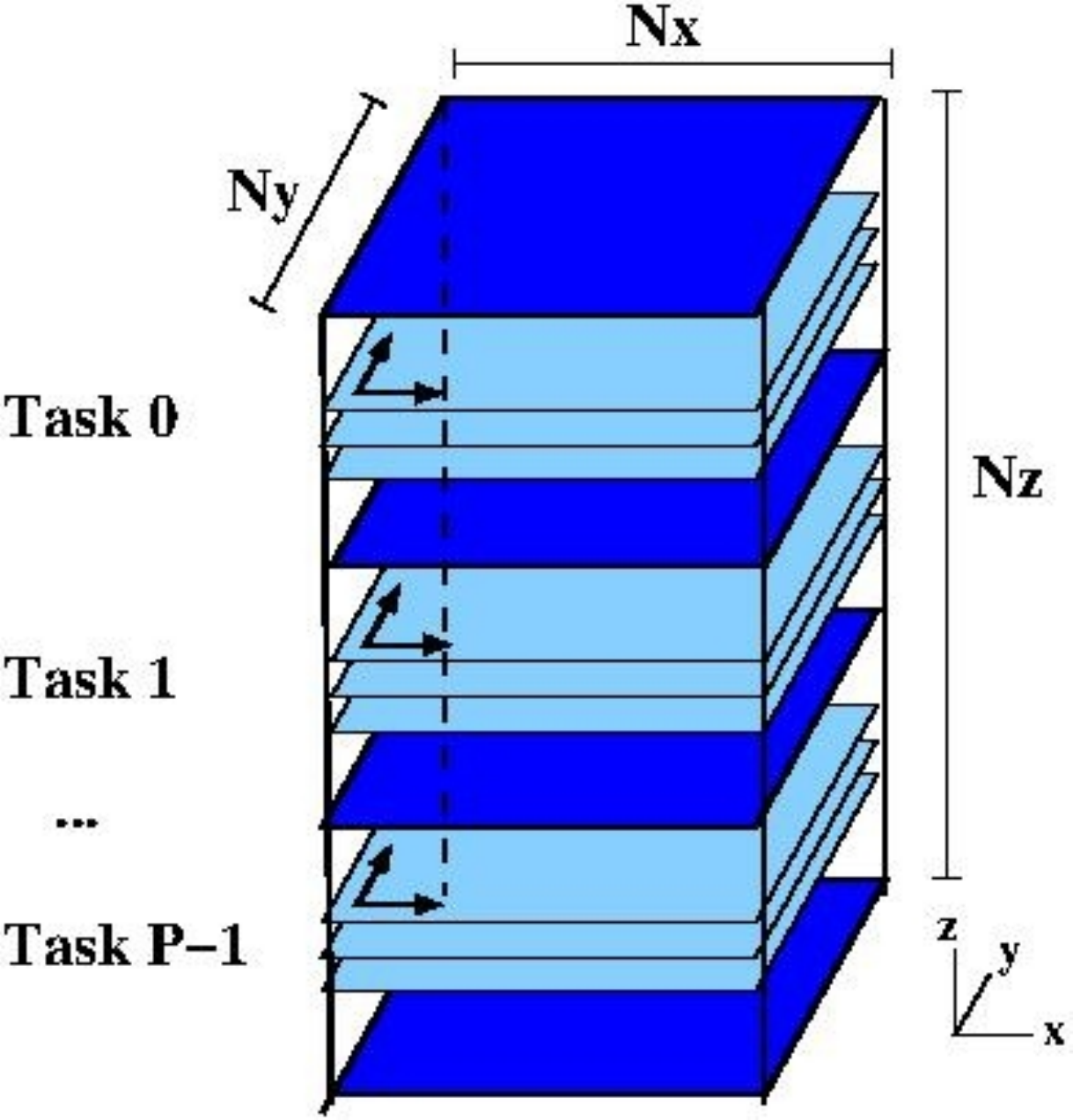} 
\includegraphics[height=20pc]{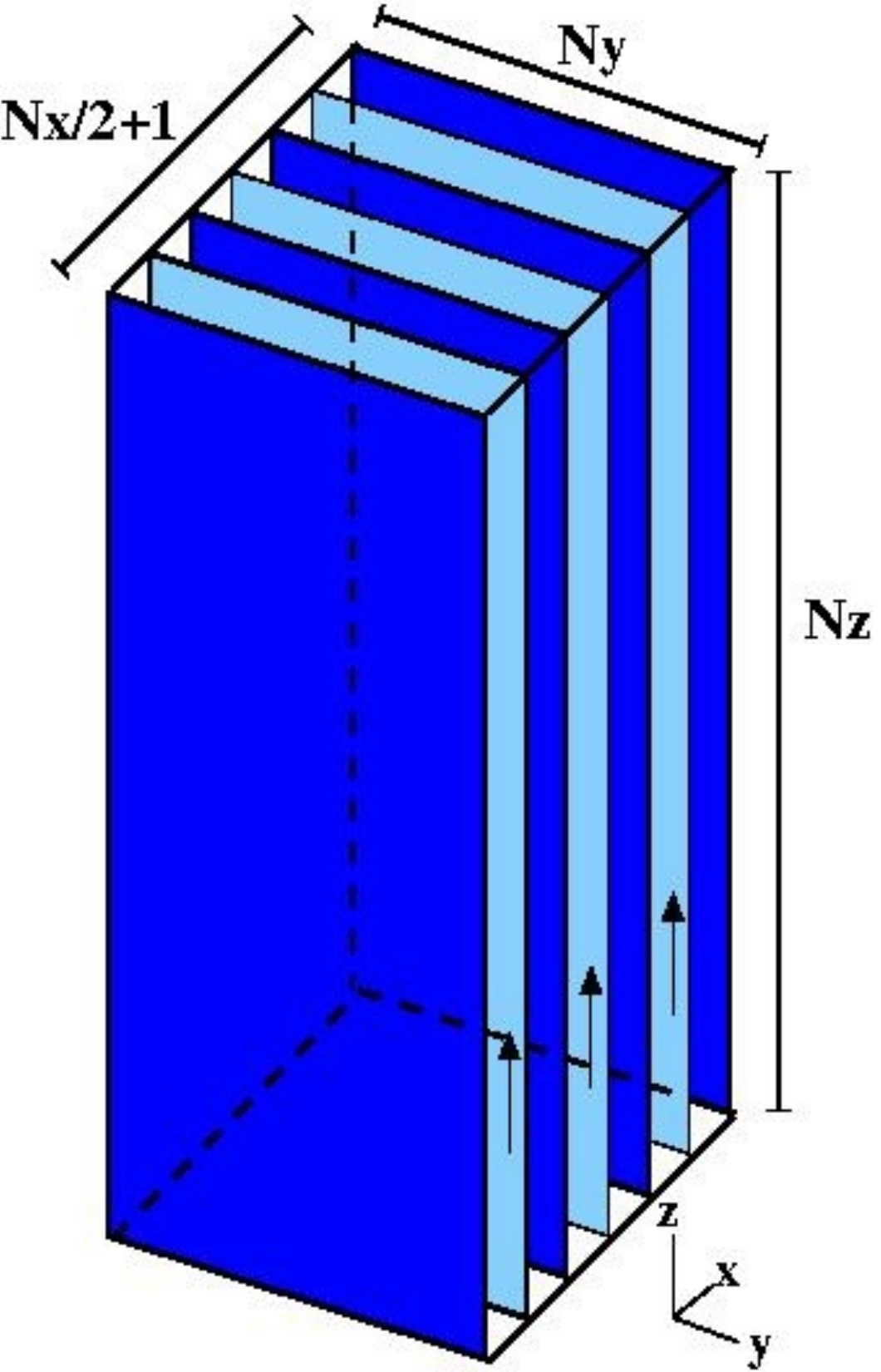} 
\end{center}
\caption{Schematic of the global grid, and its 1D or slab-based domain
  decomposition used for the global Fourier transform. {\it Left}
  shows the real data block in physical space, together with the
  decomposition of the block among MPI tasks, where the dark shaded
  planes indicate MPI task boundary. This decomposition highlights the
  $z$-incompleteness (and $x$- and $y$-completeness) of the physical
  space data. The first step of the global transform is indicated by
  the arrows and represent a local 2D transform in each plane of the
  slab owned by the task, indicated by the lightly shaded planes. 
  On the {\it right} the original data block has been transposed in a
  second global transform step, so that the partially transformed data
  o the left is now $z$- and $y$-complete (and $x$-incomplete), in
  order that the final step in the global transform may be done: a
  final 1D transform done in the direction of the arrow for the data
  in each of the complex (shaded) planes.}
\label{fig:dd}
\end{figure}

The global physical domain may be represented schematically as shown
in \Fig{fig:dd} (left). The domain is decomposed in the vertical
direction (a so-called 1D or {\it slab} decomposition) in such a way
that the the vertical planes are evenly distributed to all MPI
tasks (the slabs will be further decomposed into smaller domains using
OpenMP, as described below). A relatively common alternative to
this approach is to use a 2D ``pencil'' decomposition
\citep{yeung2005,chatterjee2018}, whose performance implications were
considered in \m11. If $P$ is the number of MPI tasks, there are
$M=N_z/P$ planes of the global domain assigned as work to each 
task, and from the figure, it is clear that each task ``owns'' a slab
of size $N_x\times  N_y\times M$ points. The forward transform (for
the moment, only considering MPI parallelization) consists of three
main steps:
\begin{enumerate}
\item 
A local 2D transform is carried out on the real physical data
in each plane (lightly shaded in the figure) of the local slab, over
the coordinates indicated by the arrows in \Fig{fig:dd} (left), which
produces a partial (complex) transform that is {\it z-incomplete}
because of the domain decomposition.

\item
Next, a global transpose (\Fig{fig:dd}, right) of the partially
transformed data block is made so that the data becomes $z$- and
$y$-complete (and, hence, $x$-incomplete).

\item
Finally, a local 1D transform is performed in the now complete
$z$-direction for each plane of the data in the partially transformed
slab (indicated by an arrow on  the lightly shaded planes in
\Fig{fig:dd}(b)), producing a fully transformed data block.
\end{enumerate}

Note that the total number of data points in \Fig{fig:dd} (left) is
now $(N_x/2+1) \times N_y \times N_z$, and reflects the fact that the
data is now complex, and that the transform satisfies the relation 
$\hat{\phi}({\bf k}) = -\hat{\phi}(-{\bf k})$ (assuming the data in
physical space is real). However, the {\it total} amount of complex
(real and imaginary) data is still the same as in the original data
block. The reverse transform essentially reverses these steps to
produce a real physical space field. On the CPU the local transforms
are computed using the Fast Fourier Transform (FFT) in the open source
FFTW package \citep{frigo1998,frigo2005}.

The global tranpose in the second step requires that all MPI tasks
communicate a portion of their data to all other tasks, in an
MPI all-to-all. This is handled using a non-blocking scheme
detailed in \citep{gomez2005,hybrid2011}, so we do not describe it
further here, except to state this MPI communication is the only
communication involved in the solution of the PDEs, and that it
usually represents a large fraction of the global transform time,  
as seen below. In addition to the communication, the data local to
each task must also be transposed, which can be identified as a
separate computational cost. Thus, the three steps involved in
computing the global FFT transform yield in turn three distinct
operations whose costs (times) will be considered further below: (1)
the local FFT (i.e., computation of the 2D and 1D FFTs), (2)
communication (to do the global transpose), and (3) the local
transpose of data.

Our previous work in \m11 demonstrated how OpenMP directives enable
loop-level thread parallelization of the algorithm described above to
compute parallel forward transforms, its implementation in the
MPI-parallelized GHOST code, and provided a detailed examination of
single NUMA node performance as well as a discussion of scaling to
large core counts. This thread parallelization, on top of the MPI
parallelization, results in operations performed by each thread over a
smaller portion of the slab that belongs to each MPI task, and thus
can be equivalent (depending on the number of MPI tasks and threads)
to a pencil decomposition, in which the MPI tasks operate over slabs,
and the threads over pencils in each slab. While the motivation in
\m11 was mainly to show the overall efficacy of this hybrid MPI-OpenMP
parallelization scheme, we also described the specific directives for
thread parallelization and the cache-blocking procedure used for the
local transpose step. Because of the centrality of the transform, we
investigate here the effect of placing as much as possible of the
multidimensional Fourier transform on GPUs using CUDA, while leaving 
other operations done by the CFD code in the CPUs. Thus, the method we
present has three layers of parallelization: MPI and OpenMP for
operations done in the CPUs, and CUDA for the operations that will we
moved to the GPUs.

In order to estimate the benefit of a potential GPU port of the
transform, we consider timings of the basic transform operations
in typical simulations at various resolutions. In \mytable{table:tfft}
are presented fractional timing results of runs solving
\eqs{eq:mom}{eq:divv} on uniform isotropic 3D grids of size
$N_x=N_y=N_z=N$ for $N=128$, $256$, $512$, and $1024$. These 
runs were made on the NCAR-Wyoming Cheyenne supercomputer 
using 4 MPI tasks per node without threading.  Each core is an Intel
Xeon E5-2697 v4 (Broadwell), and the interconnect topology is an
enhanced hypercube. No attempts were made to optimize MPI
communication via task placement or changes to the default MPI
environment. As with all subsequent timing results, the code runs in
``benchmark'' mode in which all computations are performed in solving
the PDEs--thus, both forward and backward transforms are done--but
with no intermediate I/O for $\mathcal{O}(10)$ time steps, and the
overall time as well as the operation (component) times are averaged
over this number of time steps. The table provides the fraction of
total average runtime spent on each of the transform operations
identified above.

The sum of the fractional times shows that the cumulative time
fraction of the distributed transform (counting FFTs, communication,
and transpose) is high, and reasonably constant at about 90\%, which
justifies our focus on the  distributed transform in isolation from
the remaining computations. Indeed, the last column in
\mytable{table:tfft} shows the maximum speed-up that could be achieved
if the time to do FFTs and transposition is reduced to zero (leaving
communication time the same). In terms of the component times (where
$\tim{R}{CPU}$ represents costs not included in the transform), we can
estimate this maximum possible speed-up {\it assuming no additional
  costs} by offloading to an accelerator as:
\begin{align*}
S  & \approx (\tim{Comm}{CPU} + \tim{FFT}{CPU} + \tim{Transp}{CPU} +
     \tim{R}{CPU})/(\tim{Comm}{CPU} + \tim{R}{CPU}) \\
& = 1 + (\tim{FFT}{CPU} + \tim{Transp}{CPU}) /( \tim{Comm}{CPU} +
  \tim{R}{CPU}) , 
\end{align*}
where the denominator approximates the aggregate runtime by using
acceleration on the distributed transform, assuming $\tim{FFT}{GPU}$
and $\tim{Transp}{GPU}$ both go to 0, and that $\tim{Comm}{CPU}$
does not change by simply adding the accelerators. Here,
$\tim{X}{CPU}$ indicates the time to compute the operation
$\textrm{X}$ when using CPUs only, where $\textrm{X}$ can be 
communication (Comm), local FFTs (FFT), or local transpose
(Transp). Since the aggregate time fraction of the transform component
is so high, we can neglect $\tim{R}{CPU}$. The final column in
\mytable{table:tfft} gives this potential speed-up for each
resolution.

This estimate indicates that speed-up is strongly connected to
performance of the local FFT and transform relative to the
communication time. While the potential speed-ups in the table may look
impressive, any optimization that serves to reduce the time for the
FFT and transpose on the CPU relative to the communication time will
also reduce the speed-up. An obvious optimization on the CPU is
threading (which is already implemented in the code, as discussed
above), but we will see in \Sec{sec:results} that the realized
speed-ups when using GPUs are still superior to the purely CPU-based
code even when multithreading is enabled.

\begin{table}
\caption{
Fraction of total runtime for each of the main operations involved in
the distributed transform: FFTs ($f_{\rm FFT}$), transpose ($f_{\it
  Transp}$), and communications ($f_{\rm Comm}$), for runs done with
different linear resolutions $N$ and number of cores $N_c$. The
remainder of the time in each run is spent on other computations. The 
last column gives the maximum speed-up possible if the cost for the
local FFTs and transposes are driven to zero. 
}
\label{table:tfft}
\begin{center}
\begin{tabular}{ccccccc}   
\hline
$N$ & $N_c$ & $f_{\rm FFT}$  & $f_{\rm Transp}$ & $f_{\rm Comm}$ & Max
       Speed-Up \\
\hline
128 & 16  &   0.68 &  0.08 &   0.13 & 7 \\
256 &  32 &   062 &  0.1 &   0.17 & 5 \\
512 &  64 &   0.61 &  0.09 &   0.21 & 4 \\
1024& 128 &   0.62 &  0.08 &   0.22 & 4 \\
\hline
\end{tabular}
\end{center}
\end{table}

\section{CUDA implementation}
\label{sec:implement}

The basic CUDA implementation of the distributed transform involves
interfacing directly or indirectly with CUDA code and primarily with
the CUDA runtime and cuFFT libraries. We describe in the following
subsections how this is accomplished in GHOST.

\subsection{Preliminaries}
\label{subsec:prelim}

As mentioned in \Sec{sec:setup}, our goal for the GPU implementation
of the distributed transform is to place as much work on the device
as possible for as long as possible, in order to reduce the number of
transfers. Of the three operations involved in the transform, the
communication alone will be explicitly handled on the CPU, although
future work will examine the ability of NVIDIA
GPUDirect\textsuperscript{TM} to at least reduce latency in
transferring data directly to and from the network. The other two
operations, the local transpose and local FFTs are handled with
interfaces to CUDA kernels, the former to our own CUDA transpose
kernel and the latter to the cuFFT library. GHOST is mainly a Fortran
90/95/2003 code so for interfacing directly or via C wrappers with
CUDA we rely heavily on ISO C bindings.  All calls to C or CUDA from
the GHOST code occur by way of ISO C bindings which standardize the
Fortran-to-C datatypes and also prevents us from having to maintain 
C function wrappers that accommodate compiler-specific name-mangling. 
A single ISO C binding interface is written for each routine called in
the cuFFT, the CUDA runtime library, or made to a C function that we
have written to, say, launch a CUDA kernel. An example of a binding is
given in the listing below, in which the CUDA runtime function name is
specified in the \bld{bind} clause, and the function name called from
Fortran is taken in this example to be the same:
\begin{lstlisting}
 !****************************************************************
 ! cudaHostAlloc
 !****************************************************************
 INTEGER(C_INT) function cudaHostAlloc(buffer, isize, flags) &
                bind(C,name="cudaHostAlloc")
   USE, INTRINSIC :: iso_c_binding
   IMPLICIT NONE
   TYPE(C_PTR)             :: buffer
   INTEGER(C_SIZE_T),value :: isize
   INTEGER(C_INT),value    :: flags
 END FUNCTION cudaHostAlloc
\end{lstlisting}

Many of the CUDA runtime or cuFFT library calls require parameters, or 
provide return values that may be examined, and are standardized. We
have classified these as directives (such as cuFFT\_R2C, that tells
cuFFT the direction in which to compute the transform on the GPU),
return codes both for cuFFT and the CUDA runtime library, and finally,
device properties. The cuFFT directives and return codes are taken
from the library header files, and encoded in a Fortran 90 module as
parameters. Similarly, the CUDA runtime return codes are translated
from the CUDA headers to a Fortran 'enum' type with C binding:
\begin{lstlisting}
 ENUM, BIND(C)
   ENUMERATOR ::                            &
   cudaSuccess                       =0 ,   &
   cudaErrorMissingConfiguration     =1 ,   &
   cudaErrorMemoryAllocation         =2 ,   &
   cudaErrorInitializationError      =3 ,   &
   cudaErrorLaunchFailure            =4 ,   &
   cudaErrorPriorLaunchFailure       =5 ,   &
            ...
 END ENUM
\end{lstlisting}

Finally, the CUDA device properties structure is  also taken from the
CUDA header files and made interoperable with Fortran by encapsulating 
within a Fortran structure with C binding. This type may therefore be
passed as a datatype to CUDA device property runtime calls. This
structure takes the form:
\begin{lstlisting}
 TYPE, BIND(C) :: cudaDevicePropG
   INTEGER   (C_INT) :: canMapHostMemory
   INTEGER   (C_INT) :: clockRate
   INTEGER   (C_INT) :: computeMode
   INTEGER   (C_INT) :: deviceOverlap
   INTEGER   (C_INT) :: integrated
   INTEGER   (C_INT) :: kernelExecTimeoutEnabled
           ... 
 END TYPE cudaDevicePropG
 iret = cudaHostAlloc ( plan%pccarr_, plan%szccd_, cudaHostAllocPortable) 
\end{lstlisting}

\subsection{Memory management}
\label{subsec:memory}

\begin{lstlisting}[float,caption={Partial listing of the
GFFTPLAN\_DATA distributed plan data type. This structure is contained
within the GHOST transform module. It contains the declarations for
the Fortran host pointers \bld{carr}, \bld{ccarr}, \bld{rarr}, the
device pointers used in cuFFT calls, \bld{cu\_dd}, \bld{cu\_ccd},
\bld{cu\_rd}, and the host \bld{C\_PTR} blocks, \bld{pcarr},
\bld{pccarr}, and \bld{prarr} for  passing host data to CUDA runtime
functions. The size variables \bld{sz*} give the  byte size of each
host (and device) array. CUDA stream pointers and local data sizes are
provided in \bld{pstream} and \bld{str\_*}, respectively (\cf,
\Sec{subsec:streams}). Handles for the local cuFFT plans for
real-to-complex (\bld{icuplanrc}) and  complex-to-real
(\bld{icuplancr}) transforms are also contained within this larger 
encapsulating plan.}\label{lst:fftplan}]
 USE iso_c_binding
 TYPE GFFTPLAN_DATA
    COMPLEX(KIND=GP),POINTER,DIMENSION(:,:,:) :: carr,ccarr
    REAL(KIND=GP),POINTER,DIMENSION(:,:,:) :: rarr
    TYPE(C_PTR) :: pcarr,pccarr,prarr
    TYPE(C_PTR) :: cu_cd,cu_ccd,cu_rd
    TYPE(C_PTR),DIMENSION(nstreams) :: pstream
    INTEGER(C_SIZE_T) :: szcd,szccd,szrd
    INTEGER(C_SIZE_T),DIMENSION(nstreams) :: str_szcd,str_szccd,str_szrd
    INTEGER(C_SIZE_T),DIMENSION(nstreams) :: icuplanrc,icuplancr
              ...
 END TYPE GFFTPLAN_DATA
\end{lstlisting}

The distributed transform in GHOST is a Fortran 90 module that handles
everything from set up to cleanup. The set up works in the same vein
as FFTW \citep{frigo1998,frigo2005}, by creating a 'plan' for the
distributed transform. The plan contains all data required for the
local FFTs, transposes and communication, and it also creates and
maintains the device pointers and other data required to perform local
operations on the GPU. Though in operation there are separate plans
for each of the forward and backward distributed transforms, the data
allocated to each plan is shared. It is also the responsibility of the
GHOST plan to create the cuFFT plans that are used to carry out the
local CUDA FFT transforms.

Non-device local data that are passed to CUDA runtime functions, \eg,
for copying to and from device memory, are declared within the module
as of \bld{TYPE(C\_PTR)}. An example of such declarations is provided
in Listing \Listing{lst:fftplan}. The plan establishes the sizes
required for each of these quantities, and these are determined by the
size and datatype of the input data.  There is a \bld{TYPE(C\_PTR)}
declaration for each type of data the cuFFTs need, but the same data
is used for the forward and backward transforms. They are, in all
respects, standard C pointers. Each of these \bld{TYPE(C\_PTR)}
variables is allocated on the CPU host in the creation step for a
distributed plan with a call in Fortran like
\begin{lstlisting}
 iret = cudaHostAlloc(plan%pcarr , plan%szcd , cudaHostAllocPortable) 
 iret = cudaHostAlloc(plan%ppcarr, plan%szccd, cudaHostAllocPortable) 
 iret = cudaHostAlloc(plan%prarr , plan%szrd , cudaHostAllocPortable) 
\end{lstlisting}
using the \bld{cudaHostAlloc} CUDA runtime function. This function
allocates page-locked ({\it pinned}) memory that can be accessed
directly by the GPU at higher bandwidths than, say, with
\bld{malloc}. In \Sec{subsec:streams}, we will see that this is also
required for asynchronous data transfer. In all cases, the
\bld{cudaHostAllocPortable} flag is used to specify that the memory is
pinned in all CUDA contexts. Note in this call that the \bld{C\_PTR}
type, and the data size are contained within the distributed plan (see
also \Listing{lst:fftplan}). Each \bld{cudaHostAlloc} is paired with a
call 
\begin{lstlisting}
 iret = cudaFreeHost (plan%pccarr)
\end{lstlisting}
in the clean up method of the plan, in order to free the host memory
properly.

Once this pinned host memory is allocated, the ``traditional'' Fortran
arrays can be associated with it for use in purely host-based
operations, like MPI communication, host copies, etc. The plan
therefore declares an associated Fortran {\it pointer} of the required
signature (type and rank). This Fortran pointer is associated with the
corresponding \bld{C\_PTR} block using the following ISO C binding
runtime call:
\begin{lstlisting}
 CALL c_f_pointer(plan%pcarr ,plan%carr , (/Nx/2+1,Ny,kend-ksta+1/))
 CALL c_f_pointer(plan%pccarr,plan%ccarr, (/Nz,Ny,iend-ista+1/))
 CALL c_f_pointer(plan%prarr ,plan%rarr , (/Nx,Ny,kend-ksta+1/))
\end{lstlisting}

Note that, $ksta$ ($kend$), and $ista$ ($iend$), refer to the bounding
indices of the global grid that define the work region of each MPI
task, as described in \Sec{sec:setup}: The $k$-bounds are required due
to the $z$-incompleteness of the real and partially transformed data
(\bld{rarr} and \bld{carr}), and the $i$-bounds are required because
of the $x$-incompleteness of the the full complex transform
(\bld{ccarr}) as shown in \Fig{fig:dd} (left) and \Fig{fig:dd} (right),
respectively. Thus, while the manifestly Fortran arrays are {\it not}
the central, allocated quantities used in the distributed transform
method on the host, due to the interfaces with cuFFT and CUDA runtime,
the \bld{C\_PTR} host blocks are.

Lastly, the device pointers contained within the plan are created:
\begin{lstlisting}
 iret = cudaMalloc(plan%cu_cd , plan%szcd )
 iret = cudaMalloc(plan%cu_ccd, plan%szccd)
 iret = cudaMalloc(plan%cu_rd , plan%szrd )
\end{lstlisting}
These calls, and the scope of these device pointers within the
transform module, allow for the persistence of the pointers until the
the plan's clean up method is called.

\subsection{CUDA streams for overlapping device data transfer and
  computation}
\label{subsec:streams}

\begin{figure}
\begin{center}
\includegraphics[width=14cm]{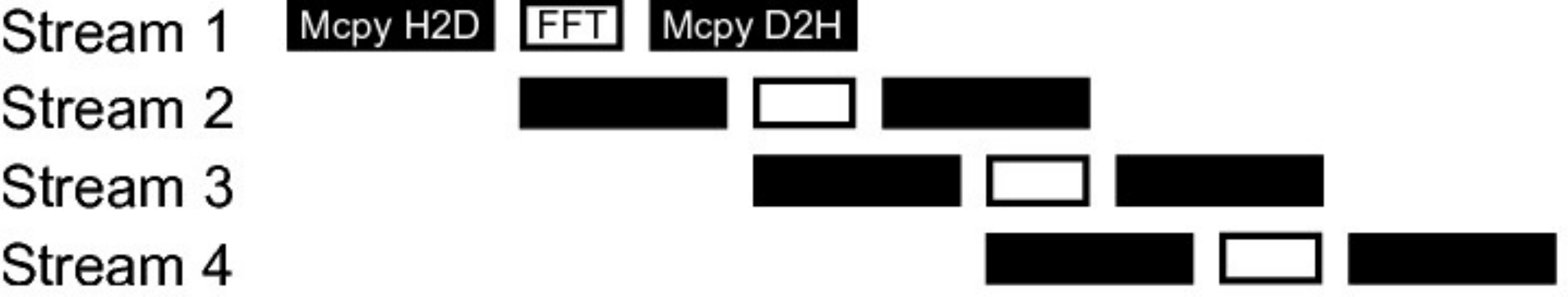} 
\end{center}
\caption{Schematic of CUDA stream use to overlap data transfer from
  the host to the device (H2D) with cuFFT computation. Each stream
  operates on only a section of the data block. The data is
  transferred asynchronously, and the computation starts on a stream
  as soon as the data transfer is complete. Black ``Mcpy H2D''
  (``D2H'') boxes indicate copies of the data from the host to the
  device (or {\it vice versa}), while white ``FFT'' boxes correspond
  to the computation of the FFTs on the data available in that
  stream.}
\label{fig:streams}
\end{figure}

CUDA streams are used in an attempt to overlap the data transfer to
the device for local cuFFT device computations
\citep{sanders2011}. The basic procedure is illustrated in
\Fig{fig:streams}, and involves three distributed transform {\it plan
  execution} steps: (1) copying the data for each stream to the device
asynchronously using  \bld{cudaMemcpyAsync}, (2) performing the local
FFT on the device using the appropriate {\it streamed} cuFFT method
for the local FFT being computed, and (3) having each stream copy the
data back to the host again using \bld{cudaMemcpyAsync}. The local
cuFFT task is completed by calling the \bld{cudaStreamSynchronize},
after which we are guaranteed that the computation has completed. Each
of  these steps is done in a ``batch'' process for the entire set of
streams by the main thread. We emphasize that the batch processing  by
the main thread highlights the fact that a single GPU is bound to a
single MPI task.

As mentioned in \Sec{subsec:memory}, the host and device storage for
each of these operations, as well as the data sizes, are computed in
the GHOST plan creation routine. The CUDA streams are also created
there using \bld{cudaStreamCreate}, and they are destroyed using a call 
to \bld{cudaStreamDestroy} in the GHOST plan clean up method. If the
local transpose operation (\cf, \Sec{sec:setup}) is performed on the
device, as it is for the results presented below, execution step (3)
must be postponed until after the transpose is complete. 

In each batch call in execution step (2), each stream calls the
appropriate cuFFT function with a cuFFT plan created for that
stream. These plans are created in the GHOST plan creation method; the
local plan handles are carried in the GFFTPLAN\_DATA in
\Listing{lst:fftplan}, where \bld{icuplanrc} are the plan handles for
real-to-complex transforms, and \bld{icuplancr} are those for
complex-to-real transforms. Each of these local stream plans uses the
cuFFT  ``advanced data layout'' for batch processing of FFTs. These
local plans specify the data stride between successive input and
output elements, and the number of input and output elements for the
FFT.  Offsets for the the input and output data in the
\bld{cudaMemcpyAsync} in steps (1) and (3) above may also be computed
in the GHOST plan creation step, and carried in the plan data for use
on demand. These same offsets can be used in the calls to the cuFFT
routines in execution step (2) to specify the starting location of the
input and output data for a stream (see \Listing{lst:planexec}). Note
that the actual data passed to the \bld{cudaMemcpyAsync} call and to
the cuFFT routines are just the host \bld{C\_PTR} blocks and the
device pointers discussed in \Sec{subsec:memory} and stored in
GFFTPLAN\_DATA. Sample code for creating the real-to-complex plans is 
provided in \Listing{lst:plancreate}. The \bld{range} routine in this
listing computes for each stream indices that are comparable to the
quantities $ksta$ ($kend$), and $ista$ ($iend$) denoting the bounds of
an MPI task's subdomain in \Sec{subsec:memory}, but in this case it
returns the beginning and end of data chunks in each stream. The value
of GFLOATBYTESZ is a constant parameter that specifies the byte size
of the configurable GHOST floating type. 

\begin{lstlisting}[float,caption={Partial code listing of the GHOST
plan creation code showing the creation of the streams and the local
stream cuFFT plans for 2D real-to-complex cuFFTs.  
}\label{lst:plancreate}]
 DO i = 1,nstreams
     iret = cudaStreamCreate(pstream(i))
 ENDDO
 nrank= 2
 DO i = 1,nstreams
    CALL range(ista,iend,nstreams,i-1,first,last)
    issta (i) = first
    issnd (i) = last
    CALL range(ksta,kend,nstreams,i-1,first,last)
    kssta (i) = first
    kssnd (i) = last
    plan%str_szccd(i) = max(2* Nz     *Ny*(issnd(i)-issta(i)+1) &
                                        *GFLOATBYTESZ,GFLOATBYTESZ)
    plan%str_szcd (i) = max(2*(Nx/2+1)*Ny*(kssnd(i)-kssta(i)+1) &
                                        *GFLOATBYTESZ,GFLOATBYTESZ)
    plan%str_szrd (i) = max(   Nx     *Ny*(kssnd(i)-kssta(i)+1) &
                                        *GFLOATBYTESZ,GFLOATBYTESZ)
    na      (2) = Nx         ; na      (1) = Ny;
    pinembed(2) = Nx         ; pinembed(1) = Ny*(kssnd(i)-kssta(i)+1);
    ponembed(2) = Nx/2+1     ; ponembed(1) = Ny*(kssnd(i)-kssta(i)+1);
    istr        = 1          ; idist       = Nx*Ny         ;
    ostr        = 1          ; odist       = Ny*(Nx/2+1)   ;
    iret = cufftPlanMany(plan%icuplanrc(i),nrank,na,pinembed,istr,idist, &
                             ponembed,ostr,odist,CUFFT_R2C,kssnd(i)-kssta(i)+1);
 ENDDO
\end{lstlisting}

A code sample of the three transform execution steps is provided in
\Listing{lst:planexec} for (most of) the forward distributed
transform. This code shows explicitly the ``batching'' of the three
execution steps using CUDA streams, as well as the use of the host and
device data previously discussed when interfacing with the CUDA
runtime and cuFFT calls. Code for the computation of the local data
transpose and for the final 1D batch cuFFT of the now $x$-complete
data is omitted, but indicated.

\begin{lstlisting}[float,caption={Partial code listing the forward
GHOST transform plan execution. The three execution steps are seen
clearly.  The byte offsets for the input and output data are given
explicitly here. Note the association of the cuFFT plans with their
corresponding streams. This must be done prior to any batch cuFFT
calls using a new plan so that the streamed cuFFT API can be used. The
\bld{cudaMemCpyAsyncOffDev2Host} is simply an ISO C-bound wrapper to
the \bld{cudaMemcpyAsync} that applies the offsets locating the
appropriate input and output data for that batch call.
} \label{lst:planexec}]
  DO i = 1,nstreams ! Associate cuFFT plans with streams 
     iret = cufftSetStream(plan%icuplanr(i),pstream(i));
  END DO
  plan%rarr = real_input_data ! Copy real input data to host pointer
  DO i = 1,nstreams ! Batch copy of input data to device
     byteoffset1 = plan%Nx*plan%Ny*(kssta(i)-ksta)*GFLOATBYTESZ
     byteoffset2 = plan%Nx*plan%Ny*(kssta(i)-ksta)*GFLOATBYTESZ
     iret = cudaMemcpyAsyncOffHost2Dev(  plan%cu_rd, & ! Dev
                                         byteoffset1, & ! OFFSET Dev
                                         plan%prarr,  & ! Host
                                         byteoffset2, & ! OFFSET Host
                       plan%str_szrd(i), pstream(i) )
  END DO
  DO i = 1,nstreams ! Batch 2D cuFFT
     byteoffset1 = plan%Nx*plan%Ny*(kssta(i)-ksta)*GFLOATBYTESZ
     byteoffset2 = 2*(plan%Nx/2+1)*plan%Ny*(kssta(i)-ksta)*GFLOATBYTESZ
     iret = cufftExecOffR2C(plan%icuplanr(i), plan%cu\_rd, & ! Dev
                                                byteoffset1, & ! OFFSET Dec
                                                plan%cu_cd, & ! Dev
                                                byteoffset2)    ! OFFSET Host
  END DO
  DO i = 1,nstreams ! Batch copy from device to host
     byteoffset1 = 2*(plan%Nx/2+1)*plan%Ny*(kssta(i)-ksta)*GFLOATBYTESZ
     byteoffset2 = 2*(plan%Nx/2+1)*plan%Ny*(kssta(i)-ksta)*GFLOATBYTESZ
     iret = cudaMemCpyAsyncOffDev2Host(  plan%pcarr, & ! Host
                                         byteoffset1, & ! OFFSET Host
                                         plan%cu_cd, & ! Dev
                                         byteoffset2, & ! OFFSET Dev
                     plan%str_szcd(i), pstream(i) )
  END DO
  DO i = 1,nstreams ! Batch synchronization
     iret = cudaStreamSynchronize(pstream(i))
  END DO
 ! <Do local transpose step>
 ! <Do final batch local 1D transform in x-complete direction>
\end{lstlisting}

Additional comments about our GPU implementation are warranted. In
general, a maximum number of streams allowed is set at build time. But
not all GPUs support overlapping data transfer with computation as we
have outlined. One should use \bld{cudaGetDeviceProperties} to check
the \bld{deviceOverlap} field of the \bld{cudaDeviceProp} structure, in
order to determine if overlapping data transfer with computation is
supported on a given device. If not, \bld{nstreams} should be set to 1
in \Listing{lst:plancreate}. Because a single MPI task is bound to a
single GPU, we cannot at this time use the multiple GPU cuFFT
transforms that are available available in cuFFT 8
\citep{cuFFT_developer_link}. Finally, it is worth pointing out that
even when CUDA is used in the distributed transform, the code not
operating on the device may still be threaded as discussed in
\m11. Given that the implementation binds an MPI task to a single
device, the extra threading is able to exploit effectively multicore
nodes that have significantly fewer GPUs than compute elements. Thus,
and as previously mentioned, the code can utilize three levels of
parallelization: a coarse (MPI) level, a finer--grain CPU threading
level (OpenMP), and a fine grain (GPU) level. All three will be in
used in \Sec{sec:results}. 

\begin{figure}
\begin{center}
\includegraphics[trim=0.6cm 6.5cm 2.4cm 6.5cm,clip=true,
  width=8cm]{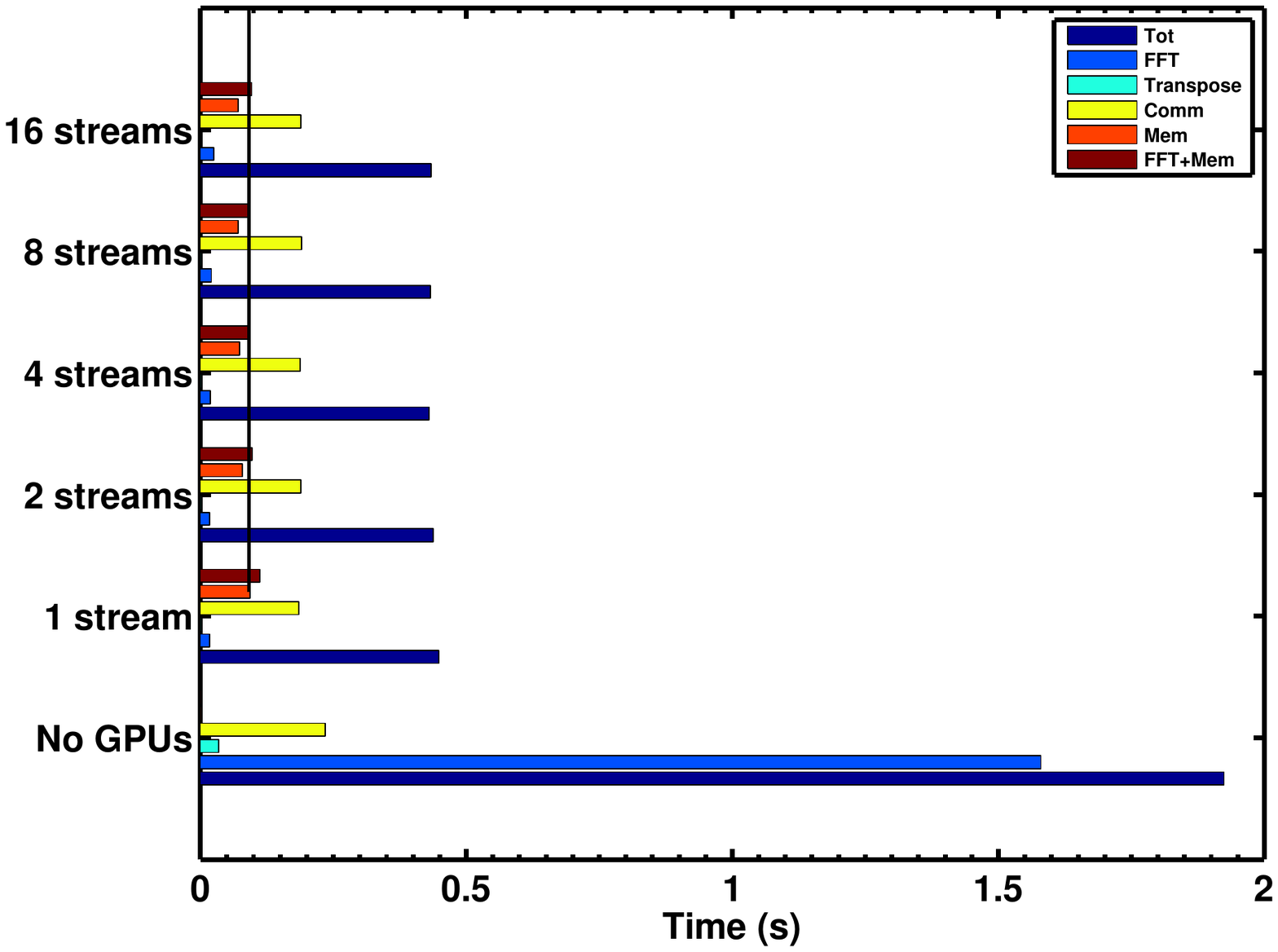}
\includegraphics[trim=0.6cm 6.5cm 2.4cm 6.5cm,clip=true,
  width=8cm]{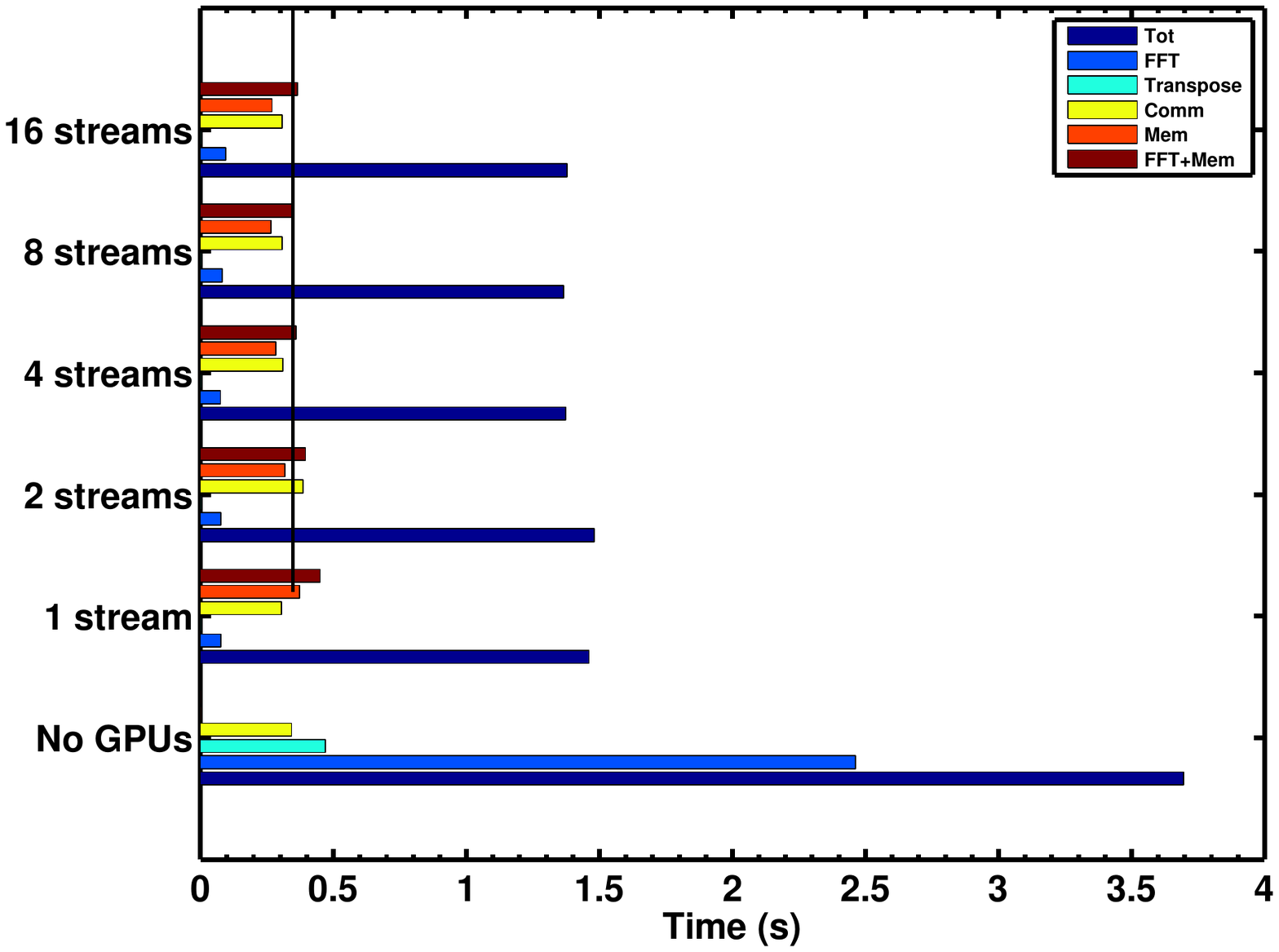}
\end{center}
\caption{Total time (Tot) and operation times for GPU runs (time to
  compute FFTs, transpose, MPI communication, memory copy from and to
  the GPU, and FFTs plus copies) with varying number of CUDA
  streams. {\it Left:} In the system with NVLink, and {\it right:} in
  the PCIe system. The minimum transfer time in both cases is seen for
  $\mathbf{nstreams} = 8$, and its time $\tim{FFT}{}+\tim{Mem}{}$ is
  indicated by the dark vertical line (compare the increase in this
  time for $\mathbf{nstreams} = 1$). In the bottom of each figure, the
  total and component times (excluding  device transfer time) are
  provided for a corresponding CPU-only run using the same number of
  MPI tasks and threads per task.}
\label{fig:stream-copy}
\end{figure}

\section{Results: scaling and performance}
\label{sec:results}

A series of tests have been performed to evaluate the implementation
discussed in \Sec{sec:implement}. Most of these tests have been
conducted on Oak Ridge Leadership Computing Facility's SummitDev
system, which contains 54 nodes each with 2 IBM Power8 chips (10 cores
each) and 4 NVIDIA Tesla P100 GPUs. The nodes utilize a full fat-tree
network via EDR InfiniBand; the GPUs are  connected by NVLink 1.0 at
80GB/s, and we use CUDA 9. For all tests using this system, we
restrict ourselves to 16 of the 20 CPU cores on each node. Using
runtime directives, the MPI tasks are placed symmetrically across the
sockets, and threads from each socket are bound to their respective
tasks; the GPUs are selected to optimize data transfer. When stated,
for comparison purposes, we have also run on the NCAR-Wyoming Caldera
analysis and visualization cluster. This system contains 30 nodes,
each containing 2 8-core Intel Xeon E5-2670 (Sandy Bridge) CPUs, and
16 of which contain two NVIDIA Tesla K20Xm GPUs using PCIe Gen2 bus,
and only CUDA 8 is available.

The test procedure is the same as described in \Sec{sec:setup}, and
the resulting timing output contains the total average runtime per
time step ($\tim{Total}{}$), and the component operation times for the
local FFTs ($\tim{FFT}{}$), for the the MPI communication
($\tim{Comm}{}$), and for the transpose time ($\tim{Transp}{}$). The
additional data transfer time ($\tim{Mem}{}$, to and from the GPUs) now
also appears. These times are measured by the main thread using
primarily the  MPI\_WTIME function, and, on the GPU, the timers wrap
the entire kernel launch and execution. We compared the results of
these times with Fortran, OpenMP, and CUDA timers, to verify
consistency. In addition to the solver for \eqs{eq:mom}{eq:divv}, 
we now also examine briefly the other solvers in \Sec{sec:setup},
since we want to evaluate the performance when different sets of PDEs
are used (requiring different  numbers of distributed transforms). As
with timings of \eqs{eq:mom}{eq:divv} presented in \Sec{sec:setup}, we
examine aggregate timings that reflect complete solutions to the
equations being considered, not just of the transform kernels. Thus,
component times are accumulated over both the forward and backward
transforms that occur in a timestep before averaging over the number
of timesteps integrated.

\begin{figure}
\begin{center}
\includegraphics[trim=0.6cm 6.5cm 2.2cm 6.5cm,clip=true,
  width=9cm]{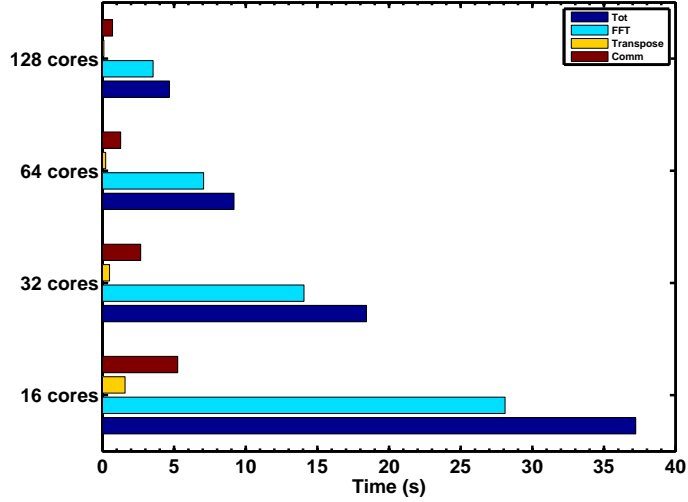} 
\end{center}
\caption{Process timers (total time and operation times) for CPU-only
  Boussinesq runs on an anisotropic grid using various core counts,
  with $\mathbf{nthd}=8$ (thus, the number of MPI tasks is equal to
  the number of cores used divided by $\mathbf{nthd}$. Timings
  correspond to the system with NVLink. Note the nearly ideal scaling
  of each of the timers with core-count.}
\label{fig:bouss-aniso-8thd-nogpu}
\end{figure}

\subsection{Variation with \bld{nstreams}}
\label{subsec:varynstreams}

We examine first the performance of the CUDA streams for overlapping
communication and computation. The pure HD equations are used here (so
the temperature  evolution equation \eq{eq:temp} is omitted, as is
the temperature term in \eq{eq:mom}). Therefore, there are seven less
distributed transforms required compared when compared to the full
system of \eqs{eq:mom}{eq:divv} (as the HD system has one less
nonlinear term when compared with the Boussinesq system). The grid is
taken to be isotropic with $256^3$ points. Sixteen cores and 2 GPUs of
one SummitDev node are used; since each GPU is bound to a single MPI
task, this implies that 2 MPI tasks each with 8 threads are
used. \Fig{fig:stream-copy} (left) shows the timings as the number of
CUDA streams is varied in this system, at the bottom the times are
compared to a CPU-only run. The first thing seen in the figure is that
there is a good speed-up over the CPU-only case, about a factor of
4.4, achieved with the best GPU time. As expected, the FFT and local
transforms times have been reduced significantly, by factors of 85 and
1300 respectively, making them nearly negligible in the GPU
runs. Unexpectedly, the communication time for the identical run
without GPUs is somewhat longer than for the GPU runs, and this
behavior is persistent for this problem.
 
Since both the local FFTs and the device transfers add up to the total
time to compute FFTs in the GPUs, we must add the two measured
times in order to determine the optimal number of streams. The
$\mathbf{nstreams} = 4$ and $8$ runs yield nearly the same aggregated
time $\tim{FFT}{}+\tim{Mem}{}$, but $\mathbf{nstreams} = 8$ gives a
slightly better result, so we adopt this as the optimal number of
streams for subsequent tests. The speed-up, $S_{\rm FFT + Mem}$, over
the time to move data to and from the GPU and to compute the FFTs, for
$\mathbf{nstreams} = 8$ over that for $\mathbf{nstreams} = 1$, 
is $21\%$. Testing on the PCIe system for comparison (see
\Fig{fig:stream-copy}, right) we also find that 
$\mathbf{nstreams} = 8$ gives the best times, but that 
$S_{\rm FFT + Mem} = 30\%$, a larger effect due to importance of using
stream optimization in the slower PCIe port. As expected, the PCIe
transfer times are uniformly slower than for NVLink (see
\Fig{fig:stream-copy}). Because of this, the improvement when using
multiple streams on the total runtime of the CFD code (when compared
with the case using GPUs with 1 stream) is about $3\%$ on the NVLink
system, versus $5\%$ on the PCIe system with the optimized streamed
transforms. Similar results were obtained when varying
$\mathbf{nstreams}$ for anisotropic grids with 
$N_x \neq N_y \neq N_z$, and for other PDEs.

\begin{figure}
\begin{center}
\includegraphics[trim=0.6cm 6.5cm 2.2cm 6.5cm,clip=true,
  width=9cm]{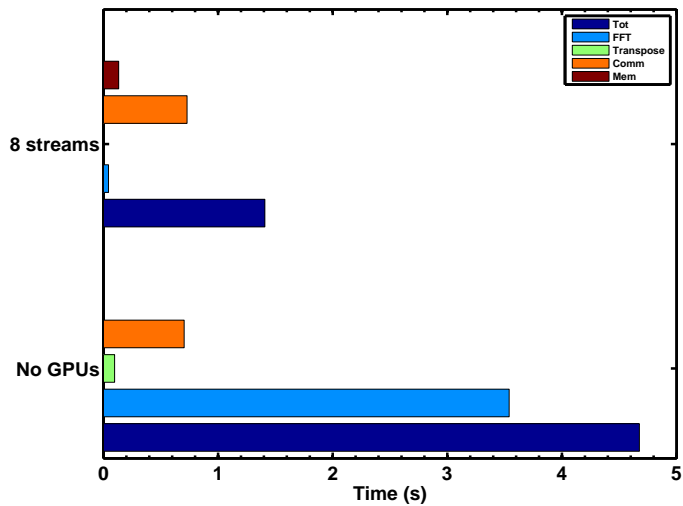}
\end{center}
\caption{Comparison between the MPI-OpenMP CPU-only code, and the
  multi-level (MPI-OpenMP-CUDA) parallel timings ($\mathbf{nstreams} =
  8$) for a Boussinesq run on a grid of $1024 \times 1024 \times 256$
  points in the system with NVLink. A total of 32 cores (2 nodes)
  were used in both cases (in the case with GPUs, this corresponds to
  8 GPUs). The overall speed-up of the GPU- to the CPU-based run is
  $3.3$.}
\label{fig:bouss-gpu-vs-cpu}
\end{figure}

Heuristically, the aggregate speed-up, $S_{\rm agg}$, afforded by the
(streamed) GPU transforms over the CPU-only version on this system may
be computed by setting the GPU times $\tim{FFT}{GPU}$ and
$\tim{Transp}{GPU}$ to zero (see \Fig{fig:stream-copy}), yielding
\begin{align}
S_{\rm agg} &= \frac{\tim{Total}{CPU}}{\tim{Total}{GPU} } \\
&\approx \frac{ \tim{Total}{CPU} }{ \tim{Comm}{CPU} + \tim{R}{CPU} +
  \tim{Mem}{}},
\label{eq:gpuspeed-up}
\end{align}
where 
$\tim{R}{CPU} = \tim{Total}{CPU} - \tim{Comm}{CPU} - \tim{FFT}{CPU}$ 
is the remainder of the time spent on the CPU that is unrelated to 
distributed transform operations. For these tests, 
$\tim{R}{CPU} \approx 0.1$. Plugging in the values 
$\tim{Comm}{CPU}\approx 0.2$, $\tim{R}{CPU}\approx 0.1$, and 
$\tim{Mem}{}\approx 0.1$ for this stream test yields 
$S_{\rm agg} \approx 4.8$, which is very close to the observed
speed-up. As expected, the speep-up results from the gains in the FFT
computed in the GPUs, minus the cost of moving the data to the device
(which is reduced by overlapping memory copies and computation using
multiple streams). From this value we also see that the speed-ups
obtained are close to the maximum we can achieve (given the fraction
of computations that were moved to the GPUs, see also
\mytable{table:tfft}), and we thus now consider the scaling of the
method with increasing number of MPI tasks and GPUs.

\subsection{Baseline CPU-only runs on anisotropic grid}
\label{subsec:cpubaseline}

As a reference, we consider first a set of CPU-only runs solving the
Boussinesq equations, \eqs{eq:mom}{eq:divv}, on an anisotropic
grid\citep{Sujovolsky2018}, using the system with NVLink. In
\Fig{fig:bouss-aniso-8thd-nogpu} we present a series of runs with the
number of threads $\mathbf{nthd}=4$ for each MPI task, using various
core counts (and, hence, number of nodes). The anisotropic domain
consists of $1024\times1024\times256$ grid points, and the figure
shows a histogram of strong scaling of each timer with core
count. Indeed, the plot indicates that each operation scales well on
this system, and that the fractional time for the distributed
transform is very similar to those in \mytable{table:tfft}. The
average parallel efficiency $\varepsilon = N_c T / (N_{c,0} T_0)$, with
respect to the reference aggregate time $T_0$ at the smallest core
count $N_{c,0}$, is about 99\% for this problem.

\begin{figure}
\begin{center}
\includegraphics[trim=0.6cm 6.5cm 2.2cm 6.5cm,clip=true,
  width=9cm]{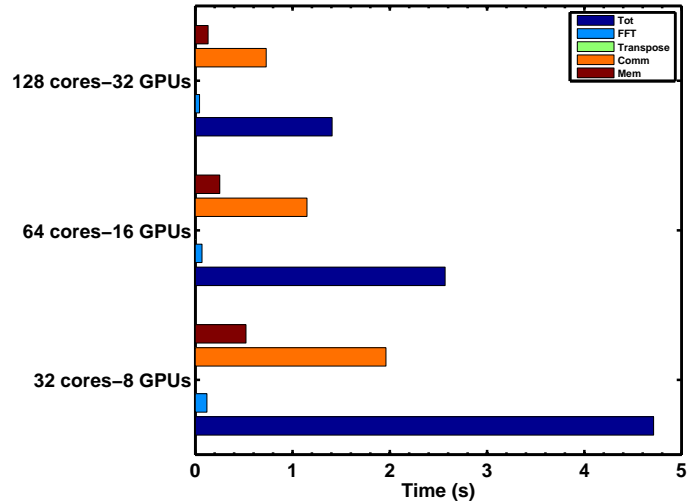}
\end{center}
\caption{Strong total and per component scaling for the problem as in
  \Fig{fig:bouss-gpu-vs-cpu}, presenting all the times measured in
  GPU-based runs varying the number of MPI tasks and of GPUs, in the
  system with NVLink. These can also be compared with the associated
  CPU-only run in \Fig{fig:bouss-gpu-vs-cpu}.}
\label{fig:bouss-gpu-scalilng-1024}
\end{figure}

\subsection{Speed-Up and scaling with full parallelization}
\label{subsec:fullparallel}

We can now consider the speed-up and scaling of the code in the
context of the Boussinesq equations with the anisotropic grid of 
$1024 \times 1024 \times 256$ points, using full MPI, OpenMP and 
GPU parallelization. For all runs in this section, we use the same
configuration of MPI tasks, GPUs and threads on each node as in
\Sec{subsec:varynstreams}, varying the number of nodes when 
required to increase the number of MPI tasks (and therefore, of
GPUs). In \Fig{fig:bouss-gpu-vs-cpu} we first present operation times
for both GPU and CPU runs on the fiducial NVLink test system using a
fixed number of 32 cores (2 nodes). We see an aggregate speed-up of
the GPU case over the CPU case of $3.3$. Like in
\Sec{subsec:varynstreams}, we observe that the FFT and local transpose
times have been reduced significantly by factors of 80 and 200,
respectively.

\begin{figure}
\begin{center}
\includegraphics[trim=0.6cm 6.5cm 2.2cm 6.5cm,clip=true,
  width=9cm]{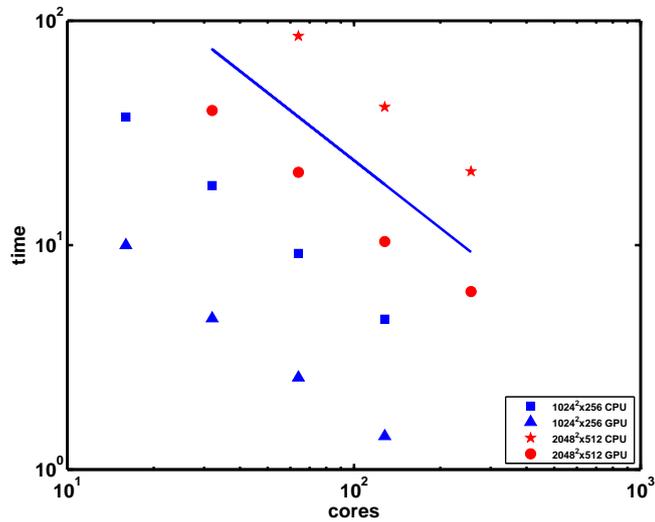}
\end{center}
\caption{Traditional strong scaling plot for anisotropic Boussinesq
  solves, comparing the CPU-only runtimes with the GPU runtimes for
  two different sets of Boussinesq runs distinguished by their
  resolutions (see legend), in the system with NVLink. Note, for fixed resolution
  and same number of cores, the significant gains when GPUs are
  used. As the figure is in log-log scale, the reference line with
  slope of $-1$ indicates ideal scaling for which the time scales as
  $1/N_c$ where $N_c$ is the number of CPU cores.}
\label{fig:bouss-gpu-scalilng-two-res}
\end{figure}

In \Fig{fig:bouss-gpu-scalilng-1024} the scaling is presented of the
same $1024 \times 1024 \times 256$ runs as in
\Fig{fig:bouss-gpu-vs-cpu}, but to larger core and GPU counts.  As
with the CPU-only results in \Fig{fig:bouss-aniso-8thd-nogpu}, the
component times and total runtimes appear to scale well with GPU count
when the GPU formulation is used. The average parallel efficiency
based on the aggregate time is slightly greater than 1 (slight super
scaling). The scaling is perhaps seen better in the more traditional
scaling plot of \Fig{fig:bouss-gpu-scalilng-two-res}, in which
CPU-only and GPU total times are provided for Boussinesq runs at two
different spatial resolutions ($1024 \times 1024 \times 256$ and
$2048 \times 2048 \times 512$ grid points). The plot is in log-log
scale, such that an ideal scaling for which the total time to perform
a time step scales as the inverse of the number of cores, $1/N_c$,
corresponds in this figure to a linear function with slope of
$-1$. The simulations for the largest resolution in
\Fig{fig:bouss-gpu-scalilng-two-res} are performed using up to 256 CPU
cores, and up to the 256 CPU cores plus 64 GPUs when the GPU-based
FFTs are used. It is clear from this figure that the scaling is nearly
ideal on this system for all cases considered. 

As a final comment, it
is worth mentioning that we conducted a production run using a preliminary
version of this code with the Boussinesq solver \citep{rosenbergbo} using
the Titan supercomputer at Oak Ridge National Laboratory. Titan has
$18688$ nodes, each with one AMD Opteron 6274 CPU (with 16 cores per
CPU) and an NVIDIA Tesla K20X GPU. Although the preliminary version of
the code used in that system did not use streams, and had local
transpositions done in the CPUs (and thus cannot be directly compared
with the results presented here, as the observed speed for that
preliminar version of the code was only of $1.7$), we still observed
good parallel scaling in a $4096^3$ run up to $10^5$ cores and $6250$
GPUs \citep[for more details, see][]{rosenbergbo}.

\begin{figure}
\begin{center}
\includegraphics[width=9cm]{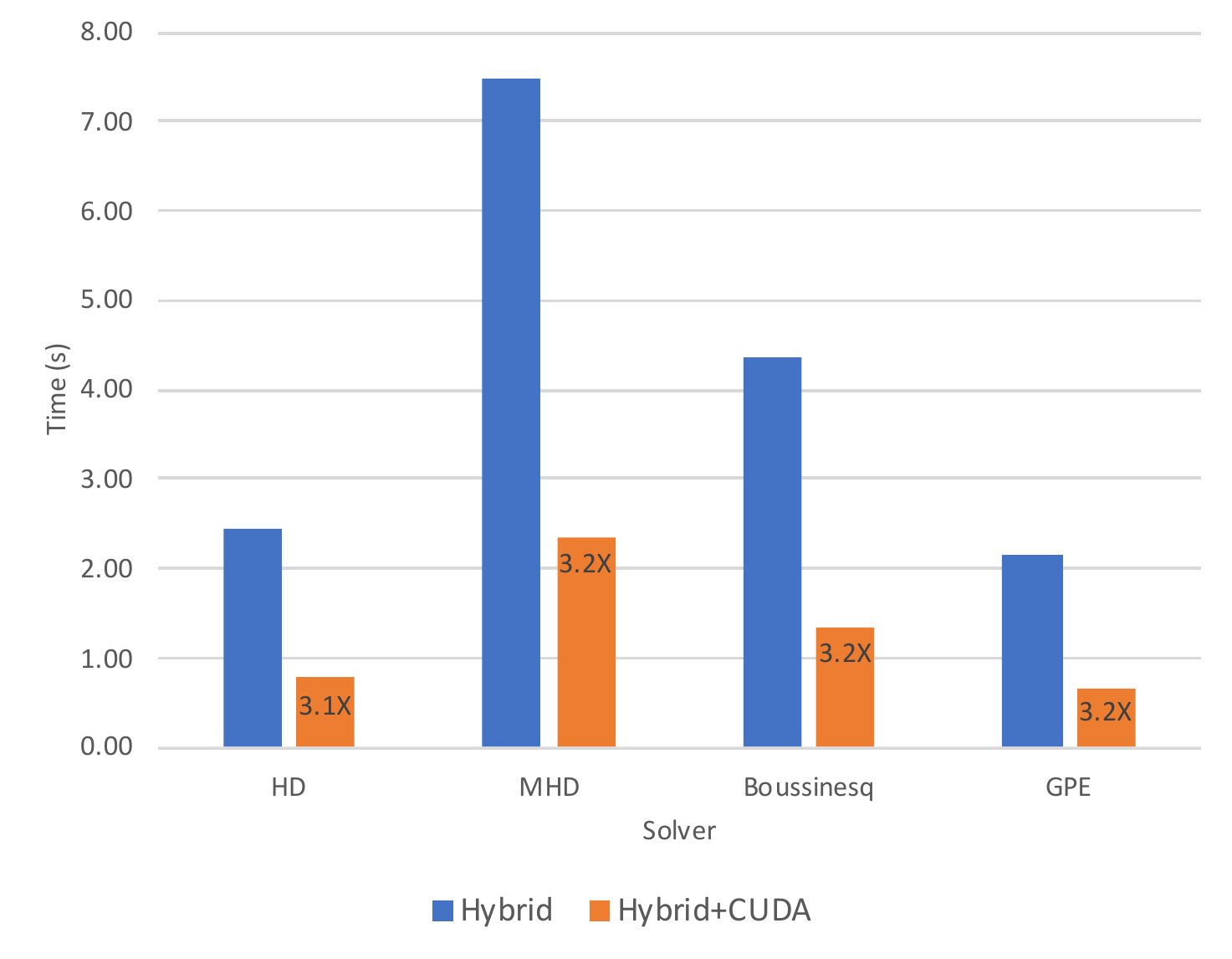}
\end{center}
\caption{Bar plot of the total time per time step (for the full CFD
  code) in CPU-only and GPU runs for each of four solvers on an
  isotropic grid of $512^3$ points with 4 nodes, 4 MPI tasks and GPUs
  per node, and 4 threads per MPI task. The speed-up of the GPU runs
  for each solver is indicated in the runtime bar for the GPU run.}
\label{fig:solver-speed-ups}
\end{figure}

\subsection{Behavior of different solvers}
\label{subsec:diffsolvers}

In the previous subsections we have shown that the the total runtime
(and component process times) for the GPU implementation follow
largely the CPU-only results in their strong scaling, when solving the
Boussinesq equations (albeit the HD equations were also briefly
considered). We expect that, as the number of nonlinear terms in the
PDEs (or the number of primitive variables) is changed by a change of 
PDEs, the overall scaling will remain good for each PDE solver for 
both the CPU-only and GPU results, and we have verified this (not
shown). We also expect that the speed-up in the runs with the CUDA
transform over the hybrid runs for different PDEs will be similar to
those in \Sec{subsec:fullparallel}, another behavior that was verified
by our tests. For the sake of brevity, here we summarize those studies
by providing some actual measurements on the NVLink test system.

\Fig{fig:solver-speed-ups} shows the aggregate runtimes for four
different solvers. Total times are provided for a CPU-only run and a
GPU run in order to compare observed speed-ups. The solvers, HD 
\citep{map2008}, MHD \citep{m1536b}, GPE \citep{Clark15a}, and 
Boussinesq \citep{rosenbergbo}, have 3, 6, 1 complex (or 2 real), and
4 primitive fields, respectively, each solver requiring a number of
distributed forward and backward transforms per time step proportional
to the number of nonlinear terms in the equations (and proportional to
the number of fields). The node layout is the same as that in
\Sec{subsec:fullparallel}, and four nodes are used for each solver,
each on a grid of $512^3$ points. The measured speed-up of the GPU
implementation of each solver over its CPU-only counterpart is
indicated. All solvers see approximately the same speed-up of 
$S_{\rm agg}=3.2$, similar to the anisotropic Boussinesq runs
above. It is clear that neither the number of distributed transforms
nor the type of grid affects the aggregate speed-up on this
system. This finding, together with the scaling (admittedly to small
node counts) gives us strong confidence that we will see a convincing
advantage in using the CUDA implementation in our production
turbulence runs on forthcoming GPU-based systems.

\section{Discussion and conclusion}
\label{sec:conclusion}

We have built upon the hybrid MPI-OpenMP parallelization scheme
presented in \m11 to add a third level of parallelization by
eveloping a CUDA implementation of the Geophysical High Order Suite
for Turbulence (GHOST) code's distributed multi-dimensional Fourier
transform. Whereas in \m11 an isotropic grid alone was considered, in 
this paper the tests admit grids with non-unit aspect ratio,
a useful device for atmospheric and ocean turbulence studies.
The method leverages the 1D coarse domain decomposition of
the basic hybrid (CPU-only) scheme, and each MPI task now binds one
GPU for additional fine grain parallelization. Our implementation
hinges on the NVIDIA cuFFT advanced data layout library for
device-capable local FFT routines, and depends considerably on the
CUDA runtime to help manage data motion between the host and
device. Implementation details have been provided that show how we
have integrated the access to the CUDA runtime and cuFFT libraries
into the code in a portable manner by using ISO bindings. We also
explain how the additional parallelism leverages the existing hybrid
(``CPU-only'') scheme, and may be useful for multicore systems that
may have fewer GPUs than cores. The resulting method is portable and
provides three levels of parallelization, allowing the usage by a
high-order CFD code of all CPU cores and GPUs in a system, and
providing typical speed-ups between 3 and 4 in tests with multiple
GPUs.

Results have been presented that test the new hybrid/CUDA, GPU-based
approach using, importantly, timings measuring the aggregate benefit
of the implementation. As mentioned, we have demonstrated that the
method can provide significant speed-up over the CPU-only computations
with the addition of GPU-based computations of the distributed
transform.  We show that the speed-ups are reasonably consistent with 
heuristics when factoring in the performance enhancement afforded by
threads for the primary costs in the transform. Aside from
communication, which remains a major issue in a distributed transform
like ours, device data transfer  times can be a significant bottleneck
with this method, but we have seen that NVIDIA's NVLink effectively
remedies this situation.

While we have seen that 80-90\% of total runtime is spent on the
distributed transform alone, the rest of the computations could in
principle be placed on the device as well. One possibility is to use
OpenACC \citep{openacc_link}, which is a directive-based programming
model that allows access to the offloading device with a rich set of
features particularly for GPUs. While this, in principle, allows us to
perform all computations on the GPU, there is a practical limit:
Turbulence codes typically do quite a large amount of diagnostic,
data-analytic, and restart output at different--but regular--intervals
during a production run. Each of these output events requires that the
CPU be engaged, so we must always be aware that data must be available
on the host. Thus, we believe our three-level method combining CPU and
GPU computations is particularly useful for massive CFD studies of
turbulent flows.

While in our implementation we transfer data from the GPU back to the
host in order to handle our all-to-all communication,
GPUDirect\textsuperscript{TM} (see \Sec{subsec:prelim}) may even help
incentivize migrating all computations to the
GPU. GPUDirect\textsuperscript{TM} enables the GPU to  share
(page-locked) memory with network devices without having to go through
the CPU using an extra copy to host memory. This should, at a minimum,
reduce communication latency, and may prove to be an important future
path.

We have concentrated on the CUDA stream implementation in this version
as a way to overlap data transfer with computation of the cuFFTs. In
general we see a benefit on runtimes of about 30\% when using PCIe,
but that this benefit is reduced when using NVLink. We have not
explored whether zero-copy access of the data by the CUDA kernels can
provide a larger gain when using PCIe. Since we are already using the
page-locked host memory, it may be possible to see still more improved
transfer speeds using  zero-copy access with the CUDA Unified Memory 
model \citep{unifiedmem_link}, but with the sizeable improvement
offered by NVLink it seems a questionable investment. On the other
hand, Unified Memory also offers the potential to clean up the data
management code presented in this work considerably. We have begun
such an implementation and will report on it in the future.

\section{Acknowledgements}
Computer time was provided by NSF under sponsorship of the National
Center for Atmospheric Research and is gratefully acknowledged. This
research also used resources of the Oak Ridge Leadership Computing
Facility, which is a DOE Office of Science User Facility supported
under Contract DE-AC05-00OR22725. Support for AP from LASP and, in 
particular, from Bob Ergun, is gratefully acknowledged.



\bibliographystyle{elsarticle-harv} 



\end{document}